\documentclass[reprint,amsmath,amssymb,aps,pra]{revtex4-2}
\usepackage{graphicx}
\usepackage{dcolumn}
\usepackage{bm}
\usepackage[hidelinks]{hyperref}
\hypersetup{colorlinks=true, urlcolor=blue, linkcolor=blue, citecolor=blue}

\begin{document}

\title{Chaos stabilizes synchronization in systems of coupled inner-ear hair cells}
\author{Justin Faber$^1$}
\email{faber@physics.ucla.edu}
\author{Hancheng Li$^2$}
\email{hanchengli@ucla.edu}
\author{Dolores Bozovic$^{1, 3}$}
\email{bozovic@physics.ucla.edu}
\affiliation{ $^1$Department of Physics \& Astronomy, $^2$Department of Electrical \& Computer Engineering, and $^3$California NanoSystems Institute, University of California, Los Angeles, California 90095, USA\\}
\date{\today}

\begin{abstract}
\noindent \textbf{Hair cells of the auditory and vestibular systems display astonishing sensitivity, frequency selectivity, and temporal resolution to external signals.  These specialized cells utilize an internal active amplifier to achieve highly sensitive mechanical detection. One of the manifestations of this active process is the occurrence of spontaneous limit-cycle motion of the hair cell bundle.  As hair bundles under \textit{in vivo} conditions are typically coupled to each other by overlying structures, we explore the role of this coupling on the dynamics of the system, using a combination of theoretical and experimental approaches. Our numerical model suggests that the presence of chaotic dynamics in the response of individual bundles enhances their ability to synchronize when coupled, resulting in significant improvement in the system's ability to detect weak signals.  This synchronization persists even for a large frequency dispersion and a large number of oscillators comprising the system.  Further, the amplitude and coherence of the active motion is not reduced upon increasing the number of oscillators.  Using artificial membranes, we impose mechanical coupling on groups of live and functional hair bundles, selected from \textit{in vitro} preparations of the sensory epithelium, allowing us to explore the role of coupling experimentally.  Consistent with the numerical simulations of the chaotic system, synchronization occurs even for large frequency dispersion and a large number of hair cells.  Further, the amplitude and coherence of the spontaneous oscillations are independent of the number of hair cells in the network.  We therefore propose that hair cells utilize their chaotic dynamics to stabilize the synchronized state and avoid the amplitude death regime, resulting in collective coherent motion that could play a role in generating spontaneous otoacoustic emissions and an enhanced ability to detect weak signals.}  
\end{abstract}

\maketitle

\section{Introduction}

The auditory and vestibular systems exhibit remarkable sensitivity, frequency selectivity and temporal resolution \cite{HUDSPETH14}.  These systems can detect vibrations that induce motion of only a few angstroms, well below the amplitude induced by thermal fluctuations in the surrounding fluid.  Humans are able to distinguish sounds that differ in frequency by only $\sim0.2\%$.  Further, we are able to resolve two stimulus impulses that differ temporally by only 10 microseconds \cite{LESHOWITZ71}.  These characteristics are crucial for identifying and localizing sounds, as well as comprehending speech, especially in noisy environments.  These phenomena, among others, are not fully understood and the physics of hearing remains an active area of research \cite{REICHENBACH14}.

Mechanical detection of auditory signals is performed by hair cells.  These specialized sensory cells are named after the rod-like stereovilli that protrude from their apical surface.  The stereovilli are arranged in interconnecting rows and are collectively named the hair bundle. Incoming sound waves or vestibular accelerations induce deflections of the hair bundle, which cause a shearing motion between neighboring stereovilli.  This shearing causes mechanically-gated ion channels to open, yielding an influx of ionic current into the hair cell \cite{LEMASURIER05, VOLLRATH07, OMAOILEIDIGH19}. The resulting changes in the membrane potential elicit further signaling from the hair cell to the auditory neurons, propagating the information that a mechanical signal has been detected.

Auditory detection has been shown to require an active energy-consuming process \cite{HUDSPETH08}. In a number of species, hair bundles have been further shown to oscillate spontaneously in the absence of external stimulus \cite{BENSER96, MARTIN03}.  These limit-cycle oscillations exhibit amplitudes significantly larger than the motion induced by the thermal fluctuations of the surrounding fluid, and they have been shown to violate the fluctuation dissipation theorem, proving them to be active \cite{MARTIN01}.  The existence and role of these spontaneous oscillations in intact animals has not yet been established.  However, the results of several studies suggest that they could be important for signal detection, as they provide a potential amplification mechanism.  Further, spontaneously oscillating hair bundles provide a probe for studying the underlying active processes of the inner ear.

Given the existence of essential nonlinearities in the auditory system, nonlinear dynamics theory has been applied to study its behavior. Specifically, the dynamics of individual hair bundles have been well described with the normal form of the Hopf bifurcation\cite{EGUILUZ00, KERN03}.  This simple differential equation reproduces many of the experimentally observed features of the hair cell dynamics, such as the sensitivity and frequency selectivity, as well as the spontaneous oscillations and the compressive nonlinear response. To achieve this extreme sensitivity and frequency selectivity, the system has been assumed to be poised at a Hopf bifurcation.  However, in the proximity of this bifurcation, the system experiences a phenomenon known as critical slowing down, meaning that a stimulus perturbing it away from the steady-state behavior will result in a long transient before returning to steady state \cite{JI18}.  This is inconsistent with the high temporal resolution of the auditory system.  To avoid the inherent trade-off between sensitivity and temporal resolution, we proposed that the system is poised deeply in the oscillatory regime, rather than in the immediate vicinity of the Hopf bifurcation.  We have previously shown that a system which exhibits chaotic dynamics in the oscillatory regime shows an enhancement of both sensitivity and rapidity of response \cite{FABER19b}.  However, in this dynamical regime, an individual uncoupled oscillator is not frequency selective.

\textit{In vivo}, hair bundles are mechanically coupled by an overlying membrane.  The nature of this coupling varies across species and across the organs of the inner ear \cite{OMAOILEIDIGH19}.  It tends to be strong and, in some cases, may suppress the spontaneous oscillations.  However, the inner ear does spontaneously emit faint tones in the absence of stimulus \cite{KEMP79}.  These spontaneous otoacoustic emissions (SOAEs) are ubiquitious across vertebrate species and occur only in live animals with intact inner ears, suggesting that they arise from an active process \cite{ROONGTHUMSKUL19}.  The mechanism responsible for their production has not yet been established, but several theoretical studies suggest they may arise from the spontaneous motion of actively oscillating coupled hair bundles, through a phenomenon known as frequency clustering \cite{VILFAN08, FRUTH14}. For actively oscillating hair bundles to produce SOAEs, they would need to overcome a phenomenon known as amplitude death, which occurs when active oscillators with large frequency dispersion are strongly coupled, resulting in quenching of the motion \cite{KIM14}.  Further, hair bundles with largely different characteristic frequencies would need to be able to synchronize in order to form the narrow spectral peaks found in SOAE recordings.

We have previously demonstrated a mechanism by which chaos can aid in an oscillator's ability to synchronize to external signals \cite{FABER19}.  In the current work, we extend this study to a system of coupled active oscillators, which provides a model for the behavior of a full auditory or vestibular end organ. Specifically, we show that this same chaotic regime causes Hopf oscillators to avoid amplitude death and instead synchronize with each other, despite large dispersion in the characteristic frequencies.  We show that this synchronization is stable, as it persists for large system sizes, providing a plausible model for biological systems.  Neither the amplitude nor the coherence of the spontaneous motion is compromised upon increasing the number of oscillators in the network.  We test these theoretical predictions by experimental studies performed on \textit{in vitro} preparations of excised epithelia, in which hair bundles were coupled using artificial membranes.  We find consistent results in our experimental studies and theoretical predictions.  Therefore, we propose that chaotic dynamics enhance the synchronization of oscillating hair bundles, causing the system to avoid the amplitude death state and instead produce spontaneous motion that could aid in signal detection, as well as result in the production of SOAEs.

Using the numerical model of this coupled system, we also demonstrate that this chaos-induced synchronization results in enhanced sensitivity and frequency selectivity to weak, external signals without compromising the speed of the response.  This mechanism provides an attractive alternative to the dynamical regime in the immediate vicinity of the Hopf bifurcation, where the system sacrifices temporal resolution due to critical slowing down.

\begin{figure*}[t!]
\includegraphics[width=\textwidth]{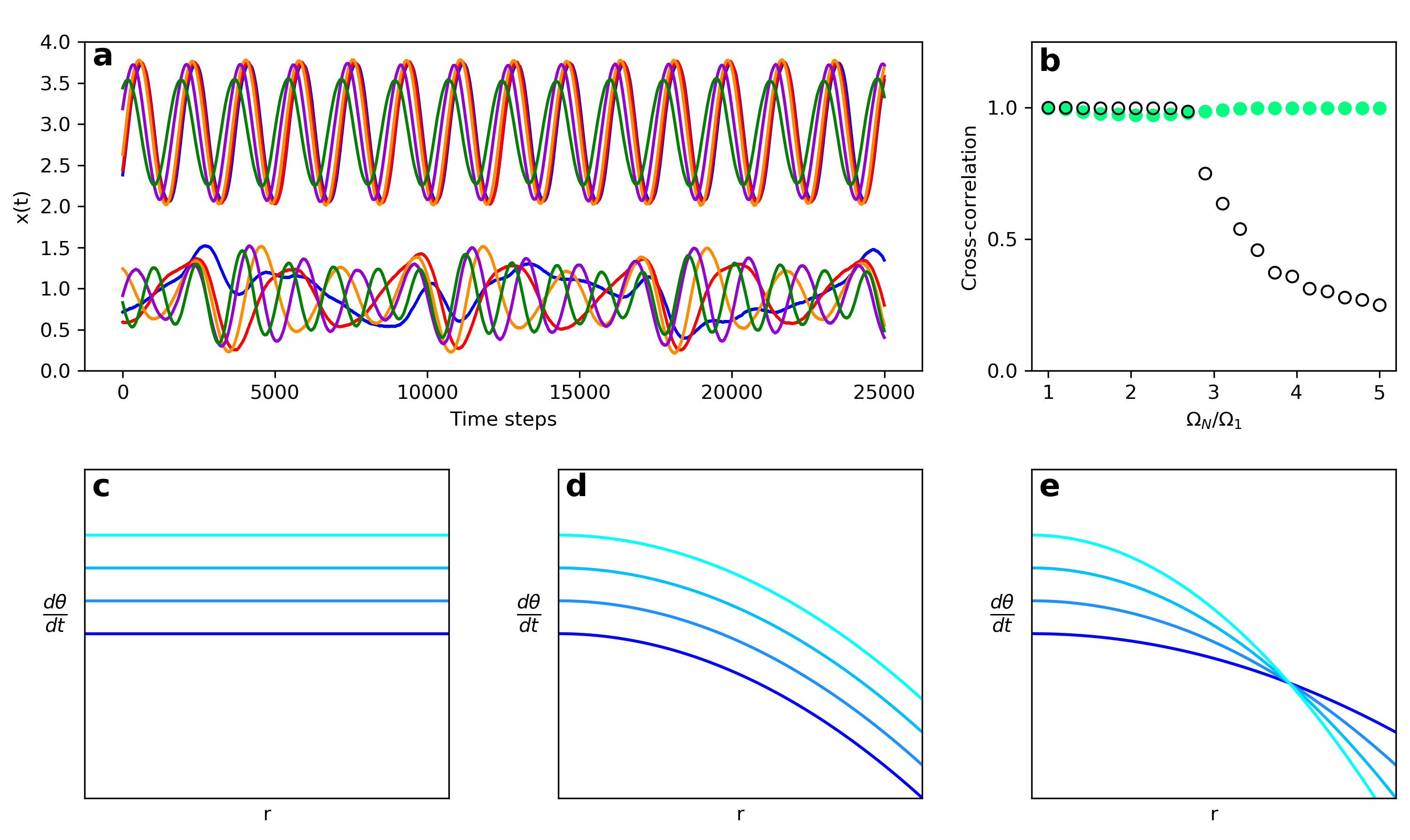}
\caption{(\textbf{a}) Time traces of 5 coupled isochronous oscillators (bottom) and 5 coupled nonisochronous oscillators with $\beta_j$ linearly spaced from 0 to 6 (top).  (\textbf{b}) Cross-correlation coefficient as a function of the frequency dispersion of a system of 5 oscillators for the isochronous (black-open) and nonisochronous (green-filled) cases.  (\textbf{c-e})  Illustrations of the instantaneous frequencies of four oscillators as a function of the oscillation amplitude for the isochronous, nonisochronous with identical $\beta_j$, and nonisochronous with dispersion in $\beta_j$ systems, respectively.}
\label{Fig1}
\end{figure*}

\begin{figure*}[t!]
\includegraphics[width=\textwidth]{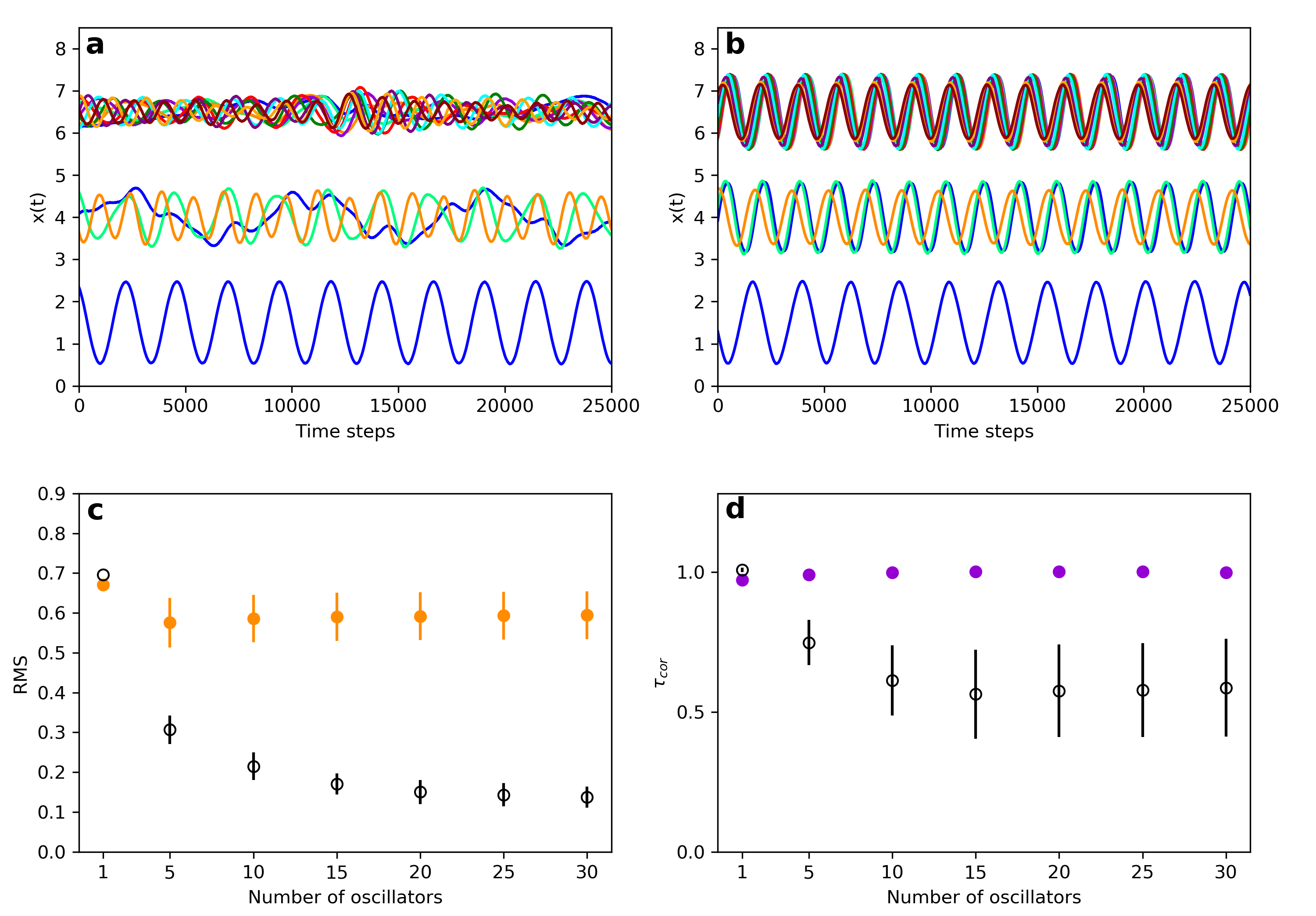}
\caption{(\textbf{a-b}) Time traces of coupled oscillators for the isochronous and nonisochronous ($\beta_{max} = 6$) systems, respectively.  The bottom, middle, and top sets of traces correspond to system sizes of $N=1$, 3, and 10, respectively.  (\textbf{c})  RMS of the autonomous oscillations for a range of system sizes for the isochronous (black-open) and nonisochronous (orange-filled) cases.  (\textbf{d}) Normalized correlation time (Eqn. \ref{tau_cor}) for the isochronous (black-open) and nonisochronous (purple-filled) systems.}
\label{Fig2}
\end{figure*}

\begin{figure*}[t!]
\includegraphics[width=\textwidth]{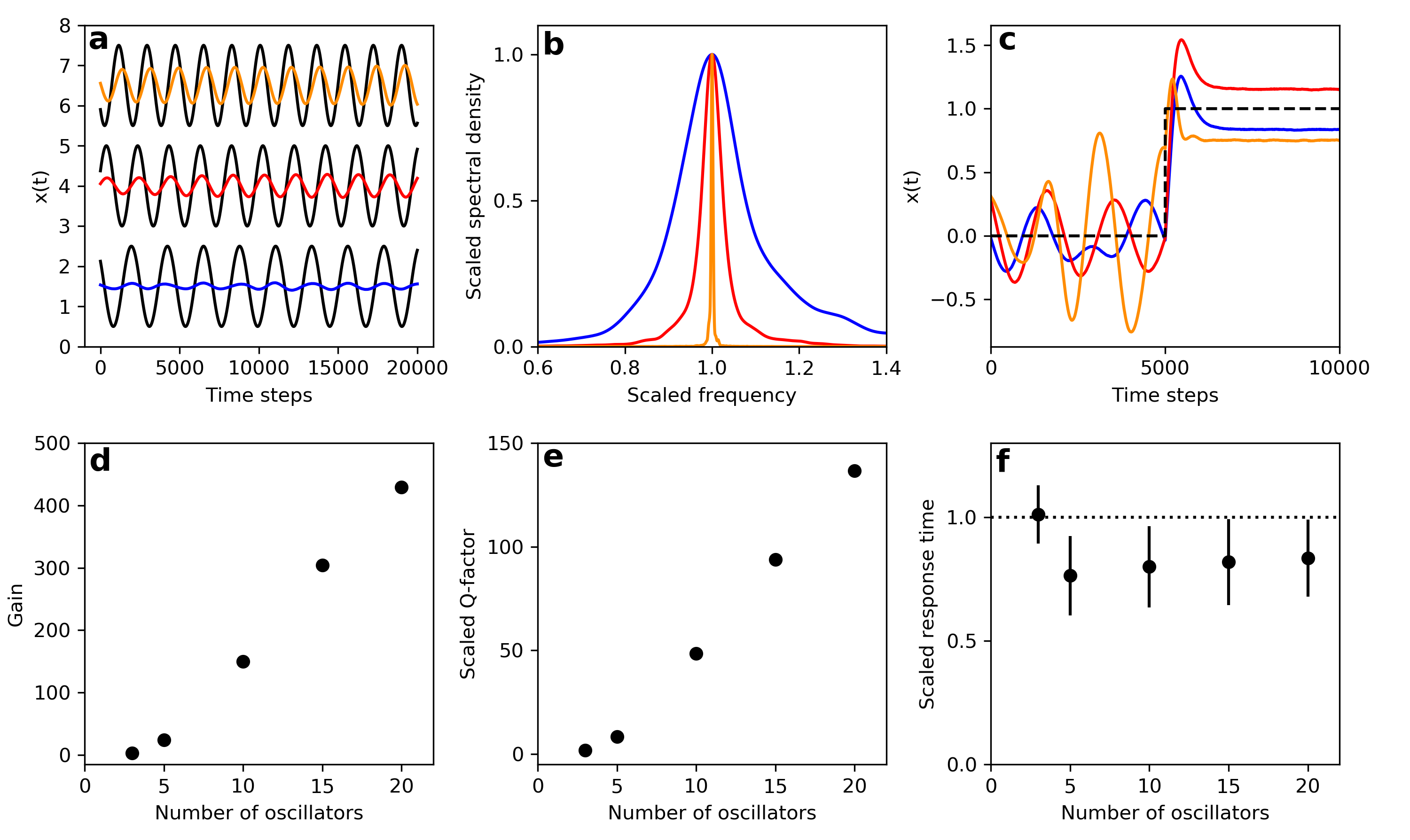}
\caption{(\textbf{a}) Response of the system to weak sinusoidal stimulus, indicated by the black curves.  (\textbf{b}) Spectral curves in response to low-level white-noise stimulus.  (\textbf{c}) Average oscillator response to a step stimulus, as indicated by the black-dashed curve.  For (a-c), the system size was $N=10$, and the blue, red, and orange curves represent the $\beta_{max} = 0$ (isochronous), $\beta_{max} = 2$, and $\beta_{max} = 6$ systems, respectively.  (\textbf{d}) Spectral value of the nonisochronous ($\beta_{max} = 6$) system at the resonance frequency in response to weak sinusoidal stimulus, scaled to the spectral response of the isochronous system.  (\textbf{e})  Quality factor of the system with $\beta_{max} = 6$ in response to weak white-noise stimulus, scaled to the quality factor of the isochronous system.  (\textbf{e}) Response time of the nonisochronous ($\beta_{max} = 6$) system to a step stimulus, scaled to the response time of the isochronous system.  Points and error bars represent the mean and the standard deviation for 100 presentations of the stimulus.}
\label{Fig3}
\end{figure*}

\section{Numerical Model of \\ Coupled Hair Bundle Dynamics}

The dynamics of the $j^{th}$ oscillator in the system are governed by the normal form equation for the supercritical Hopf bifurcation, with complex additive white Gaussian noise:

\begin{eqnarray}
\frac{dz_j(t)}{dt} = (\mu + i\omega_j)z_j(t) - (\alpha + i\beta_j)|z_j(t)|^2z_j(t)\nonumber   \\  + k\Big(S(t) - x_j(t)\Big) + \eta_j(t),
\label{eq:Hopf}
\end{eqnarray}

\noindent where

\begin{eqnarray}
z_j(t) = x_j(t) + iy_j(t).
\end{eqnarray}

\noindent Here, $x_j(t)$ represents the bundle position, while $y_j(t)$ reflects internal parameters of the bundle and is not assigned a specific measurable quantity.  However, the existence of this hidden variable is essential to reproduce the experimentally observed dynamics.  $\mu$ represent the control parameter of the oscillators, which determines the proximity to the Hopf bifurcation.  The natural frequency at this bifurcation point is given by $\omega_j$. For an individual, uncoupled oscillator, the limit-cycle radius is given by $r_0 = \sqrt{\frac{\mu}{\alpha}}$, and the limit-cycle frequency at finite radius is $\Omega_j = \omega_j - \beta_j r_0^2$.

All oscillators are coupled to the overlying artificial membrane with coupling stiffness, $k$.  The position of the membrane, $S(t)$, is governed by the differential equation,

\begin{eqnarray}
m\frac{d^2S(t)}{dt^2}  + \lambda\frac{dS(t)}{dt} = \sum_{j=1}^N k \Big(x_j(t) - S(t)\Big) + \eta_s(t),
\end{eqnarray}

\noindent where $m$ and $\lambda$ represent the mass and drag of the artificial membrane, respectively.  The membrane is subject to white Gaussian noise represented by the real-valued stochastic variable, $\eta_s(t)$.  The dynamics of the individual oscillators are represented by complex variables and are hence subject to complex additive white Gaussian noise with independent real and imaginary parts.  All noise terms in the model are independent, with correlation functions

\begin{eqnarray}
\langle \eta_a(t)\eta_a(t') \rangle = 2D \delta (t-t')
\end{eqnarray}

\noindent and

\begin{eqnarray}
\langle \eta_a(t)\eta_b(t') \rangle  = 0,
\end{eqnarray}

\noindent for $a \neq b$.  Here, $D$ represents the noise strength, which we set to be identical for all noise terms.

$\alpha$ and $\beta_j$ characterize the nonlinear term of the system.  In most prior studies, $\beta_j$ was set to zero, rendering the oscillators isochronous.  For such a system, the frequency is independent of the amplitude of oscillation.  However, when $\beta_j \neq 0$, the system is nonisochronous, and the instantaneous frequency depends on the amplitude of the limit cycle.  This results in more complex behavior and causes the additive noise to induce chaotic dynamics in the individual oscillators \cite{FABER19b}.

Hair bundle dynamics occur at a Reynolds number much below one \cite{CIGANOVIC19}.  This allows us to ignore the inertial forces of the artificial membrane ($m=0$).  Since the drag of the membrane is fairly small in comparison to the drag of the hair bundles (see supplemental material), we choose $\lambda = 0.1$.  We set $\mu = \alpha = 1$, poising the system far from the Hopf bifurcation.  We use a significant coupling stiffness of $k=2$ and a low level of noise $D=10^{-4}$, unless otherwise stated.  We vary $\beta_j$, $\Omega_j$, and $\omega_j$ throughout this study and define the limit-cycle frequencies of the slowest and fastest oscillators in a system of $N$ oscillators to be $\Omega_1$ and $\Omega_N$, respectively.  The other oscillators have limit-cycle frequencies uniformly spaced between $\Omega_1$ and $\Omega_N$.  All numerical simulations were performed using the fourth-order Runge-Kutta method with time steps of $10^{-3}$.

\section{Theoretical Results}

A nonisochronous system can modify its oscillation frequency by adjusting its amplitude, thus allowing it to easily entrain to off-resonant frequencies. As a result, two coupled oscillators with large frequency dispersion can synchronize.  Further, if the degree of nonisochronicity of the oscillators differs in correspondance with the dispersion of characteristic frequencies, syncronization can be greatly enhanced in systems of many oscillators.  In Fig. \ref{Fig1}c-e we illustrate this effect by plotting the instantaneous angular frequency, $\frac{d\theta}{dt}$, as a function of the radius of the oscillations, $r$.  We plot these curves for four oscillators with frequency dispersion and show that the curves intersect when we include dispersion in $\beta_j$.  Oscillators tend to meet at or near the intersection points, with synchronization enhanced even if the curves do not all intersect at the same point.  We perform simulations of the numerical model and compare the isochronous case ($\beta_j = 0$) to the nonisochronous systems, where $\beta_j$ varies linearly between 0 and $\beta_{max}$, in accordance with $\omega_j$.  We set $\Omega_1 = 1$ and $\Omega_N = 2\sqrt{5} \approx 4.47$, choosing the values to be the similar to the frequency dispersion observed in the experiments.  Sample traces of these simulations are plotted in Fig. \ref{Fig1}a.  We simultaneously modify $\beta_j$ and $\omega_j$ to adjust the level of nonisochronicity, while keeping the limit-cycle frequencies, $\Omega_j$, fixed.

We assess the stability of the synchronized state of five coupled Hopf oscillators as a function of the frequency dispersion.  In the isochronous case, synchronization becomes unstable for large frequency dispersions, pushing the system into the incoherent state.  Upon an increase in the coupling strength, the isochronous system transitions into the amplitude death regime, and the system becomes quiescent.  However, in the nonisochronous system, the synchronized state persists even with 5-fold frequency dispersion (Fig. \ref{Fig1}b).  Further, for the nonisochronous system, the stability of the synchronized state preserves the amplitude and coherence of the oscillators, rendering these measures independent of the system size (Fig. \ref{Fig2}). This is in contrast to the isochronous system, for which  the oscillation amplitude and coherence fall off with increasing network size.

We next determine the effects of nonisochronicity on the system's ability to detect weak signals.  We apply a weak Gaussian white noise stimulus, with noise strength, $D=0.01$, to all of the hair bundles ($x_j$) and calculate the power spectrum of the response of the oscillator $x_{(N+1)/2}$, which displays the median natural frequency. This method assumes the noise strength to be small enough to warrant consideration of only the linear response of the system. 

The nonisochronous system exhibits much higher sensitivity and simultaneously provides a more narrow band-pass filter on the white-noise stimulus in comparison to the isochronous system (Fig. \ref{Fig3}a-b).  We quantify the increase in sensitivity by finding the maximum value in the power spectrum and normalizing it by the maximum value of the power spectrum of the isochronous system.  This measure of gain indicates the factor by which nonisochronicity enhances the sensitivity of the system (Fig. \ref{Fig3}d).  Likewise, we calculate the quality factor of these peaks and normalize them by the quality factor of the isochronous system (Fig. \ref{Fig3}e).  We find that these measures of sensitivity and frequency selectivity increase with system size, consistent with prior theoretical studies \cite{DIERKES12}.  For a system of 20 oscillators, the synchronization induced by nonisochronicity leads to a sensitivity increase of over 400-fold and a frequency selectivity increase of over 100-fold.

Lastly, we show that this large enhancement in the sensitivity and frequency selectivity of response does not come at the cost of reduced temporal resolution, in contrast with close proximity to a Hopf bifurcation.  We provide an abrupt step-function stimulus to the system and average the responses of all of the oscillators.  We then calculate the time it takes for the averaged response to settle to a constant value.  As the plateau value fluctuates due to the additive noise, we calculate the time required to settle within 5 standard deviations from the mean plateau value.  We use this method to characterize the response time or temporal resolution of the system.  We scale the response time of the nonisochronous system to that of the isochronous system and show that nonisochronicity not only does not degrade the temporal resolution, but in fact slightly enhances the rapidity of the response.  Further, the speed of the system is independent of the system size (Fig. \ref{Fig3}c, f).

\section{Experimental Methods}

\subsection{Biological Preparation}
Experiments were performed \textit{in vitro} on hair cells of the American bullfrog (\textit{Rana catesbeiana}) sacculus, an organ responsible for detecting low-frequency air-borne and ground-borne vibrations.  Sacculi were excised from the inner ear of the animal, and mounted in a two-compartment chamber with artificial perilymph and endolymph solutions \cite{BENSER96}.  Hair bundles were accessed after digestion and removal of the overlying otolithic membrane \cite{MARTIN01}.  All protocols for animal care and euthanasia were approved by the UCLA Chancellor's Animal Research Committee in accordance with federal and state regulations.  

\subsection{Artificial Membranes}
Mica powder was added to a vial of artificial endolymph solution.  This solution was thoroughly mixed and then filtered through several steel mesh gratings.  These gratings served as band-pass filters to separate the mica flakes into several desired sizes.  This process was expedited by using vacuum suction to pull the solution through the grating.  The solution containing the artificial membranes was pipetted into the artificial endolymph solution, above the biological preparation.  Many of the membranes would land in the desired orientation and adhere to hair bundles underneath.  These hair bundles could then be imaged through the transparent artificial membranes.

\subsection{Data Collection}
Hair bundle motion was recorded with a high-speed camera at framerates between 250 Hz and 1 kHz.  The records were analyzed in MATLAB, using a center-of-pixel-intensity technique to determine the position of the center of the hair bundle in each frame.  The motion was tracked along the axis of channel opening/closing.  Typical noise floors of this technique, combined with stochastic fluctuations of the bundle position in the fluid, were 3--5 nm.

\begin{figure*}[t]
\includegraphics[width=\textwidth]{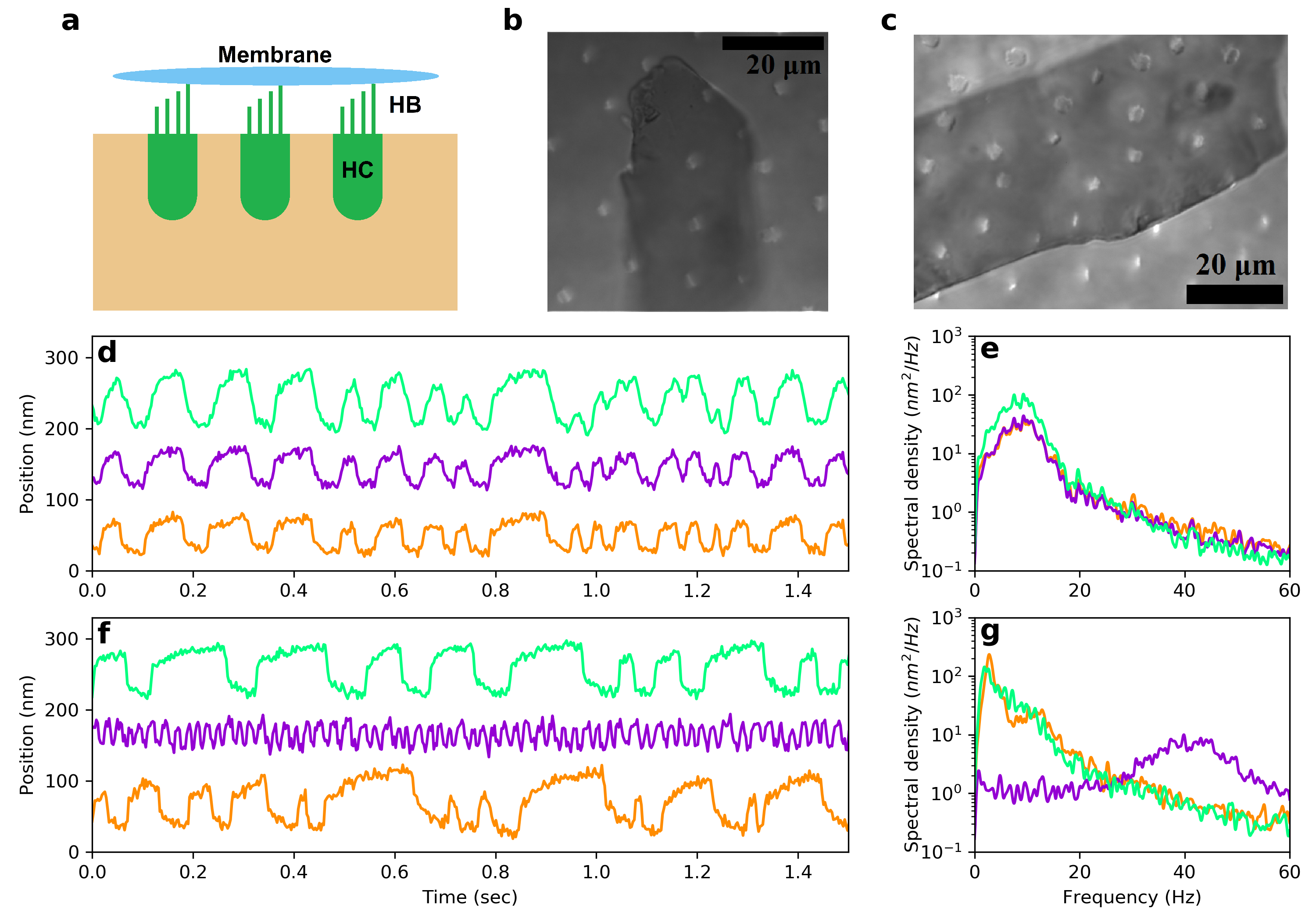}
\caption{(\textbf{a}) Illustration of the experimental system from a side view, displaying the hair cells (HC), hair bundles (HB), and an artificial membrane.  (\textbf{b-c}) Top-down images of biological preparations.  The hair bundles appear as white dots, and the shadow cast by the transparent artificial membrane can be seen in the center of the images.  (\textbf{d-e}) Time traces and power spectra of 3 spontaneously oscillating hair bundles coupled by an artificial membrane.  (\textbf{f-g}) Time traces and power spectra of the 3 hair bundles in (d-e), after removal of the artificial membrane.}
\label{Fig4}
\end{figure*}

\begin{figure*}[t]
\includegraphics[width=\textwidth]{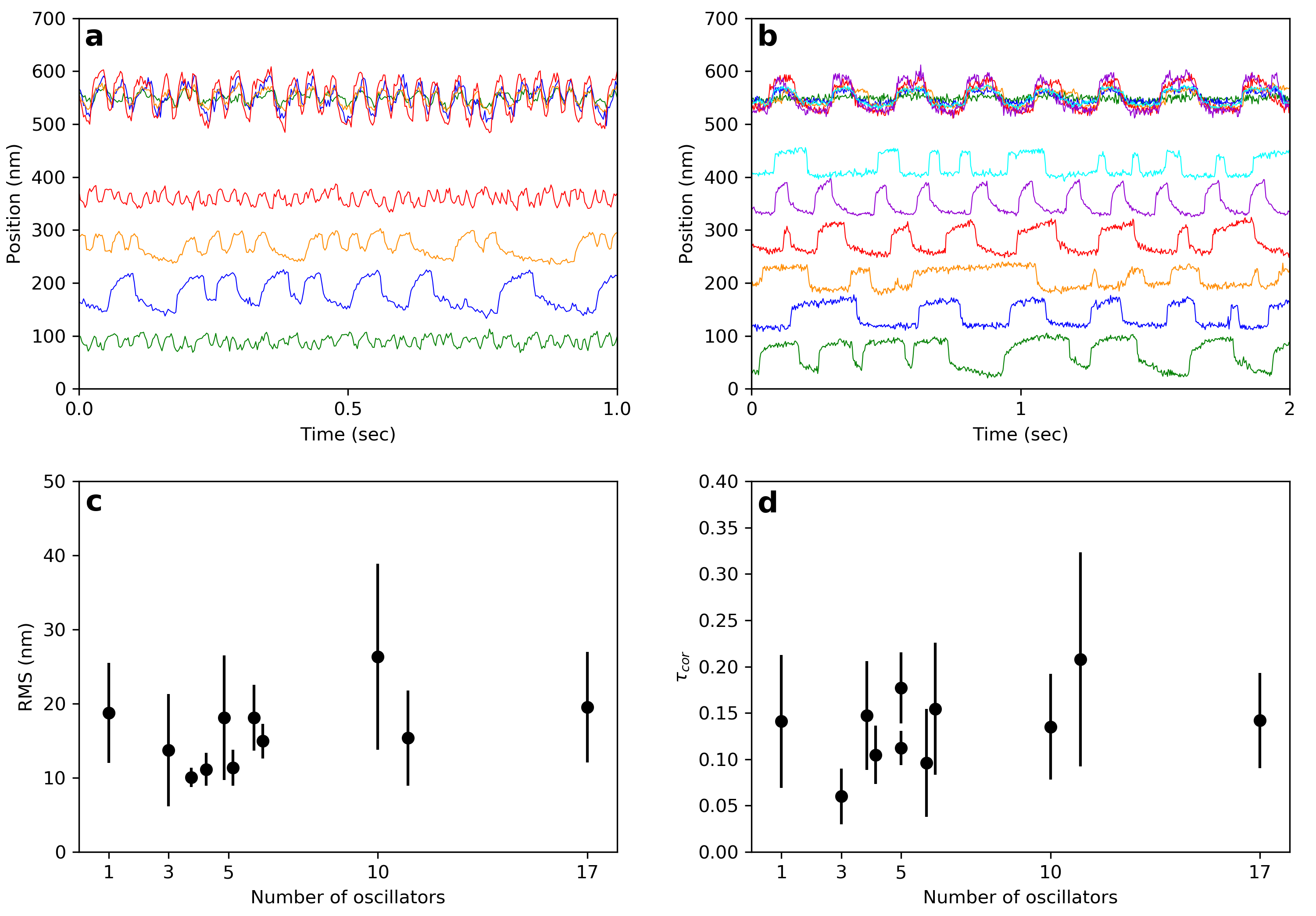}
\caption{(\textbf{a-b}) Overlaid traces of coupled hair bundles (top) for system sizes of $N=4$ and $N=6$, respectively.  Below the overlaid traces are the traces of the individual hair bundles obtained in the absence of coupling.  (\textbf{c-d}) Root mean square (RMS) amplitude and the normalized correlation time of spontaneous oscillations of coupled hair bundles, obtained for various system sizes.  Each hair bundle had a cross-correlation coefficient of at least 0.1 with other bundles in the network.  For both panels, points and error bars represent respectively the mean and the standard deviation of the coupled oscillators in the system.  For a system with $N>1$, each data point represents a separate experiment.  The points and error bars at $N=1$ represent the collective mean and standard deviation, obtained across all the experiments in the absence of coupling.}
\label{Fig5}
\end{figure*}

\section{Experimental Results}
   
To experimentally test our theoretical predictions, we created hybrid systems, in which groups of biological hair cells were artificially coupled by mica flakes of various sizes (see Experimental Methods). The mica membranes were introduced into the solutions bathing the top surface of the biological epithelia, and allowed to adhere to the underlying hair bundles, thus providing coupling. As the thin sheets of mica are transparent, they allow for precise imaging of the motion of the underlying hair bundles (Fig. \ref{Fig4}a-c). Hair bundles often exhibited synchronization, despite dispersion in their natural frequencies as large as 5-fold (Fig. \ref{Fig4}d-g) \cite{ZHANG15}, consistent with our theoretical predictions for nonisochronous oscillators.

We characterize synchronization between spontaneously oscillating hair bundles using the cross-correlation coefficient

\begin{eqnarray}
C\big(x_1(t), x_2(t)\big) = \frac{\langle \tilde{x}_1(t) \tilde{x}_2(t) \rangle}{\sigma_1 \sigma_2},
\end{eqnarray}

\noindent where $\tilde{x}_1(t) = x_1(t) - \langle x_1(t) \rangle$ and $\tilde{x}_2(t) = x_2(t) - \langle x_2(t) \rangle$ represent the time traces of the motion, $\sigma_1$ and $\sigma_2$ represent their respective standard deviations, and the angled brackets denote the time average.  $C=1$ indicates perfectly correlated motion, while $C \approx 0$ is indicative of completely uncorrelated motion.  We find the noise floor on this measure by calculating $C$ between 1225 unique pairs of uncoupled hair bundles.  The histogram of these cross-correlation coefficients has a standard deviation of approximately 0.02, with no points exceeding 0.1 (see supplemental material, Fig. S1).  To consider a pair of hair bundles to be coupled, we define our threshold to be $C \geq 0.1$, which is five standard deviations above the mean.

We compare the amplitude (root mean square) of the hair bundles' spontaneous oscillations across different sizes of artificial membranes, and hence, different sizes of coupled networks.  Due to the variation in heights of neighboring hair bundles, not every bundle under the membrane makes contact with it or becomes coupled.  Therefore, we define a network by considering only those hair bundles that have motion correlated to another bundle in the network.  As described above, we use a cross-correlation threshold of 0.1 to ensure that every bundle in the network is coupled (Fig. \ref{Fig5}a, c).  As an additional test, we repeat the calculation for a higher cross-correlation threshold of 0.5, ensuring that all of the oscillators in the network are synchronized (see supplemental material, Fig. S2).  In both cases, we consistently observe that the amplitude of synchronized motion is not reduced with increasing number of coupled hair bundles. This finding  is consistent with our theoretical predictions and indicates that hair bundles behave as nonisochronous oscillators.

We also measure the coherence of the spontaneous oscillations, which can be characterized by integrating the squared autocorrelation function to compute the correlation time \cite{PIKOVSKY97},

\begin{eqnarray}
T_{c}\big( x(t) \big) = \int_{0}^{\infty} \bigg( \frac{\langle \tilde{x}(t) \tilde{x}(t+t') \rangle}{\sigma^2} \bigg)^2 dt'.
\end{eqnarray}

\noindent Due to the finite length of the experimental recordings, we truncate the integration at two mean periods of the spontaneous oscillations.  We choose this duration, as the oscillations in the autocorrelation function have typically decayed after two full periods, and further integration would introduce unnecessary noise into the measure.  Further, we scale this measure to the correlation time of a sine wave:

\begin{eqnarray}
\tau_{cor} = \frac{T_{c}\big( x(t) \big)}{T_{c}\big( \sin(t) \big)}.
\label{tau_cor}
\end{eqnarray}

\noindent Therefore, perfectly sinusoidal motion yields $\tau_{cor} = 1$, while white Gaussian noise yields $\tau_{cor} \approx 0$.

We compare the coherence across network sizes of all coupled oscillators (Fig. \ref{Fig5}b, d) and of just those displaying synchronization (see supplemental material, Fig. S2).  Consistent with the theoretical predictions for coupled nonisochronous oscillators, the coherence does not fall off upon increasing the number of oscillators in the network.

\section{Discussion}

Auditory and vestibular systems have provided an experimental testing ground for concepts in nonequilibrium thermodynamics \cite{DINIS12}, condensed matter theory \cite{RISLER04}, and nonlinear dynamics \cite{EGUILUZ00}.  How active hair cells exhibit notable performance as signal detectors, displaying sensitivity of response, frequency selectivity, and high temporal resolution, all within a noisy fluid environment, is a long-standing open question in this area of study.  Further, auditory organs tend to contain overlying structures that impose a strong degree of coupling between individual hair cells, which in turn exhibit a large degree of dispersion of the characteristic frequencies.  It has not been established which role the presence of both strong coupling and significant frequency dispersion play in achieving the detection characteristics, or how the system avoids amplitude death to form clusters of synchronized oscillators necessary for generating SOAEs.  

Simulations of our numerical model of coupled hair bundles indicate that the nonisochronicity of the oscillators, which results in chaotic dynamics, is responsible for this robust synchronization.  The synchronization yields great enhancement of the system's sensitivity and frequency selectivity to weak external signals.  Unlike proximity to the Hopf bifurcation, this enhancement does not come at the cost of reduced temporal resolution.  Further, this synchronization persists for large numbers of oscillators and despite large frequency dispersion.  Neither the amplitude nor the coherence of the oscillations are reduced upon increasing the number of oscillators.  These results are consistent with the remarkable signal-detection attributes of the auditory system and the experimental observations of sharp spectral peaks in the SOAE recordings.

The results from our experimental recordings of coupled hair bundles are consistent with those of the numerical model.  By coupling various numbers of hair cells with artificial membranes, we find that hair bundles with differences in characteristic frequencies as large as 5-fold still routinely synchronize.  Further, the amplitude and coherence of the spontaneous oscillations are both independent of the number of hair bundles in the network.  These results can be reproduced by the numerical model only when the oscillators are chaotic ($\beta \neq 0$).  This suggests that the instabilities that give rise to chaotic dynamics of the individual hair bundles enhance the synchronization and the signal detection of the coupled system.

Chaos is often considered a harmful element in dynamical systems and something to be avoided.  For example, a chaotic heartbeat is an indicator of cardiac fibrillation \cite{GARFINKEL97}.  However, it has also been established that chaotic oscillators can easily synchronize with each other or entrain to an external signal \cite{GARFINKEL92, NEIMAN11}, as instabilities that give rise to chaotic dynamics can make the oscillators more adaptable to modifications in their autonomous motion.  Since biological systems tend to have many degrees of freedom and contain nonlinearities, chaos may be a ubiquitous element in their dynamics.  We speculate that chaos may be important in other biological systems where timing, sensitivity, and synchronization are desired, especially sensory systems responsible for detection of external signals.

\section*{Acknowledgments}
The authors gratefully acknowledge support of NSF Biomechanics and Mechanobiology, under grant 1916136. The authors thank Dr. Sebastiaan Meenderink for developing the software used for tracking hair bundle movement.


\bibliography{Bibliography}
\end{document}


\title{Online Supplemental Material for: \\ Chaos Stabilizes Synchronization in Systems of Coupled Inner-Ear Hair Cells}
\author{Justin Faber}
\author{Dolores Bozovic}
\maketitle

\section{Approximation of Artificial Membrane Drag}

We approximate the artificial membranes as infinitely thin circular disks.  Due to the low Reynolds number of hair bundle dynamics, we assume the system obeys Stokes' law.  The Stokes' drag of an infinitely thin circular disk moving edgewise through an infinite fluid is given by \cite{Trahan06}

\begin{eqnarray}
\lambda_s = \frac{16\eta d}{3},
\end{eqnarray}

\noindent where $\eta$ is the dynamic viscosity of the fluid and $d$ is the diameter of the disk.  We use the viscosity of water, $\eta \approx 10^{-3} Pa \cdot s $, and our approximate experimental range of membrane diameters ($20-50 \mu m$).  The diameters of the membranes are significantly larger than their thicknesses ($<1 \mu m$), so we consider the infinitely-thin disk to be a reasonable approximation.  Further, the boundary of the fluid is $\sim 1 cm$ away from the structures of interest, much farther than the length scale of the membranes, so the assumption of an infinite fluid is reasonable.  

We, therefore, approximate the drag coefficients of the artificial membranes to be $\lambda_s \approx 100 - 250 nN \cdot s \cdot m^{-1}$.  We compare this value to the drag coefficient of individual free-standing hair bundles.  It has previously been shown that most of the drag contribution of hair bundles comes from the channel-gating friction \cite{Bormuth2014}.  The lower bound on the total drag coefficient of an individual free-standing hair bundle was estimated to be $\lambda_0 = 425 \pm 70 nN \cdot s \cdot m^{-1}$.  Therefore, in or numerical simulations, we use a small value for the membrane drag ($\lambda = 0.1$), as it contributes only minimally to the drag of the entire coupled system.

\newpage

\section{Cross-Correlation Coefficient Noise Floor}

\begin{figure}[h!]
\includegraphics[width=\columnwidth]{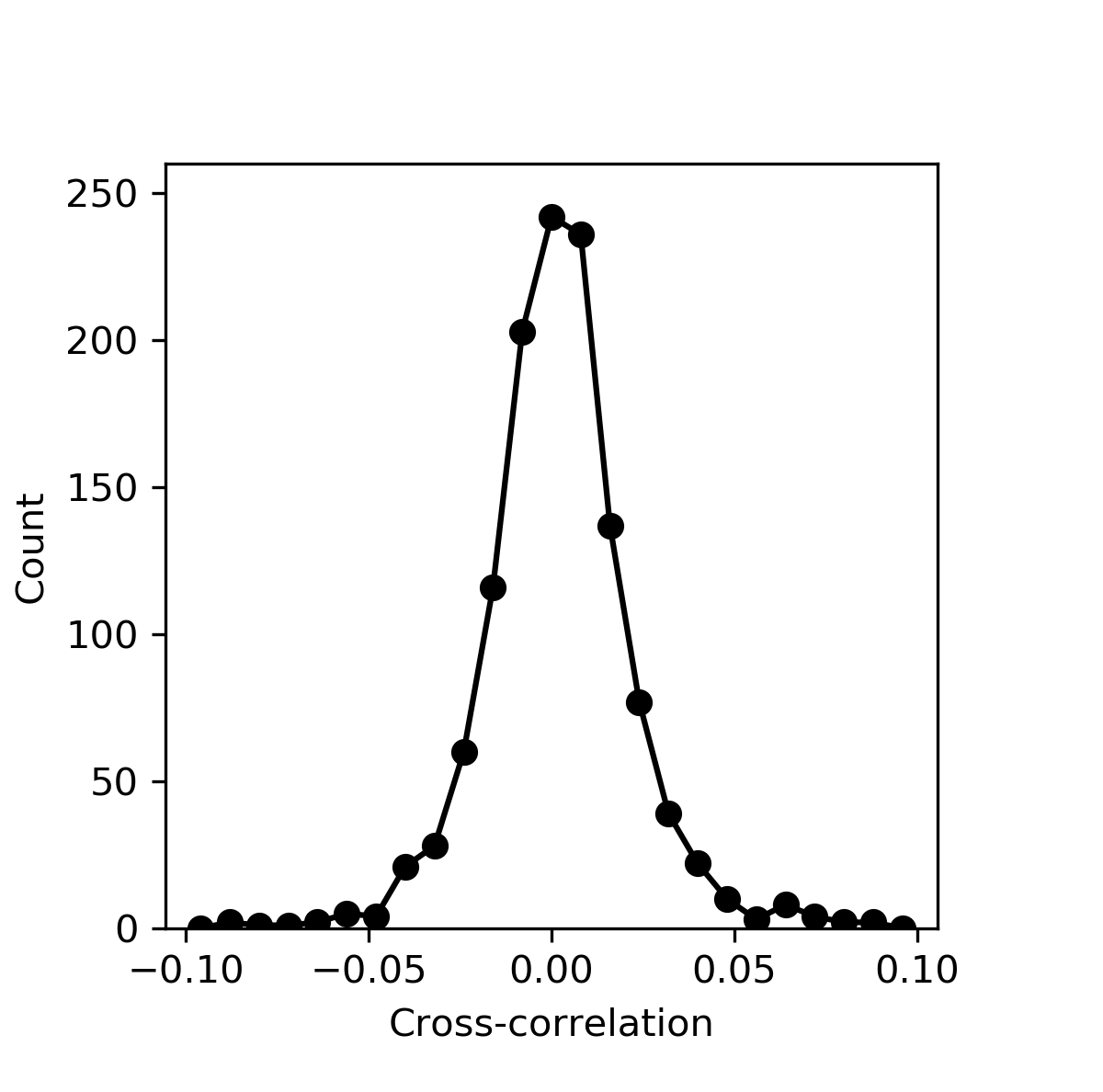}
\caption{Histogram of the cross-correlation coefficients between pairs of uncoupled, spontaneously oscillating hair bundles (1225 unique pairs).  The standard deviation of this distribution is $<0.2$ and no points exceed $0.1$.}
\label{FigS1}
\end{figure}

\clearpage

\section{Amplitude and Coherence of Synchronized Hair Bundles}

\begin{figure}[h!]
\includegraphics[width=\columnwidth]{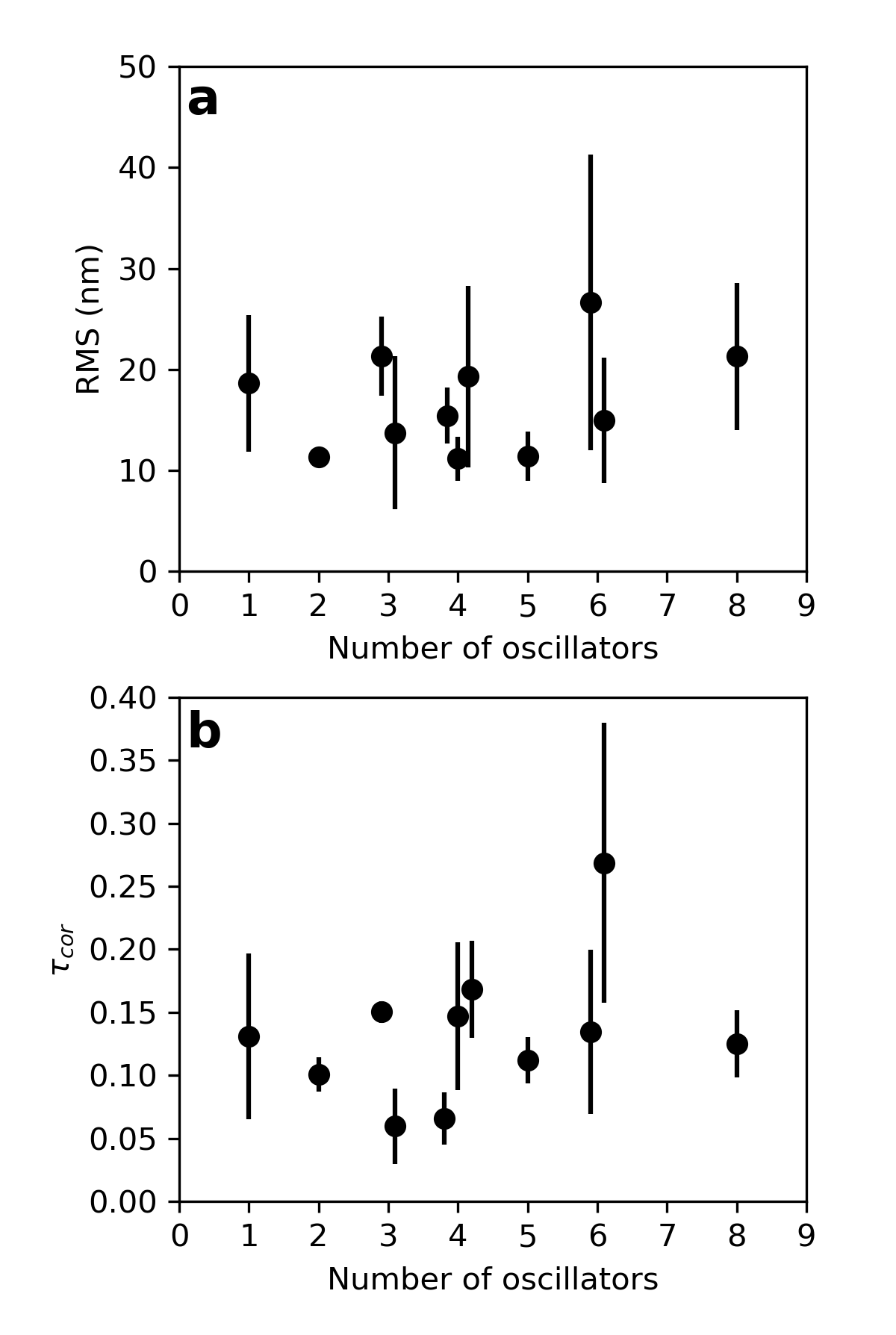}
\caption{(\textbf{a-b}) Root mean square (RMS) amplitude and the normalized correlation time of spontaneous oscillations of coupled hair bundles, obtained for various system sizes.  Each hair bundle had a cross-correlation coefficient of at least 0.5 with other bundles in the network in order to ensure the network is synchronized.  For both panels, points and error bars represent respectively the mean and the standard deviation of the coupled oscillators in the system.  For a system with $N>1$, each data point represents a separate experiment.  The points and error bars at $N=1$ represent the collective mean and standard deviation, obtained across all the experiments in the absence of coupling.}
\label{FigS2}
\end{figure}

\bibliography{Bibliography}

\clearpage